\documentclass[twocolumn,showpacs,%
amsmath,amssymb]{revtex4}

\usepackage{subfigure}
\usepackage{amsmath,amssymb}
\usepackage{graphicx}
\usepackage{rotating}
\usepackage{bm}
\usepackage{color}

\definecolor{blue}{rgb}{0,0,1}

\definecolor{green}{rgb}{0,1,0}

\definecolor{red}{rgb}{1,0,0}

\begin{document}

\author{Holger Gies}
\author{Klaus Klingm\"uller}
\affiliation{Institut f\"ur Theoretische Physik, Philosophenweg 16, %
   69120 Heidelberg, Germany} 

\title{Casimir edge effects}

\begin{abstract}
  
  We compute Casimir forces in open geometries with edges, involving
  parallel as well as perpendicular semi-infinite plates. We focus on
  Casimir configurations which are governed by a unique dimensional
  scaling law with a universal coefficient. With the aid of worldline
  numerics, we determine this coefficient for various geometries for
  the case of scalar-field fluctuations with Dirichlet boundary
  conditions. Our results facilitate an estimate of the systematic
  error induced by the edges of finite plates, for instance, in a
  standard parallel-plate experiment. The Casimir edge effects for
  this case can be reformulated as an increase of the effective area
  of the configuration.

\end{abstract}


\maketitle

\section{Introduction}

Casimir's prediction for the force $F$ per unit area $A$ between two
perfectly conducting infinite parallel plates at a distance $a$
\cite{Casimir:dh},
\begin{equation}
\frac{F_\|}{A}=-2 \gamma_\|\, \frac{\hbar c}{a^4},\quad 
\gamma_\|= \frac{\pi^2}{480}\simeq2.056\times10^{-2},
\label{eq:Casimir}
\end{equation}
has a remarkable property: a straightforward dimensional analysis
already fixes the powers of $\hbar$, $c$, and $a$ uniquely. In absence
of any other dimensionful quantity, the effects of quantum
fluctuations in this geometry can be summarized by a simple number: $2
\gamma_\|$. This coefficient is universal in the sense that it does
not depend on the microscopic details of the interactions between the
fluctuating field and the constituents of the surfaces. It is
completely fixed by specifying the geometry, the nature of the
fluctuating field and the type of boundary conditions. For instance,
for a fluctuating real scalar field with Dirichlet boundary
conditions, the parallel-plate coefficient reduces exactly to
$\gamma_\|$; the factor of 2 in Eq.~\eqref{eq:Casimir} can be traced
back to the two polarization modes of the electromagnetic field.

Away from the ideal Casimir limit, corrections to
Eq.~\eqref{eq:Casimir} arise from finite conductivity, surface
roughness, thermal fluctuations and deviations from the ideal
geometry. All these come with additional dimensionful scales, such as
plasma frequency, length scales of roughness variation, temperature or
surface-curvature radii. The corrections generically cannot be
predicted from dimensional analysis, but its functional dependence on
the further parameters has to be computed \cite{Klimtchiskaya:1999,%
Lambrecht:1999vd,Sernelius,Bezerra:2000,Bordag:2001qi,Milton:2001yy,%
Lambrecht:2005}.

The present work is devoted to an investigation of the Casimir force
between disconnected rigid surfaces, which exhibits properties similar
to Casimir's classic parallel-plate configuration: unique dimensional
scale dependencies and universal coefficients. The first property
implies that the geometry is characterized by only one length scale,
such as the distance parameter $a$. New Casimir configurations
therefore necessarily involve edges, whose influence on the Casimir
effect is an interesting and difficult question in itself. In view of
the rapid progress in the fabrication and use of micro- and nano-scale
mechanical devices accompanied by precision measurements of the
Casimir forces in these systems
\cite{Lamoreaux:1996wh,Mohideen:1998iz,Roy:1999dx,Ederth:2000,%
Chan:2001,Chen:2002,Decca:2003td}, a detailed understanding of Casimir
edge effects is indispensable.

Straightforward computations of Casimir edge effects are conceptually
complicated, since the fluctuation spectrum carries the relevant
information in a subtle manner. A technique that facilitates Casimir
computations from first field-theoretic principles is given by {\em
  worldline numerics} \cite{Gies:2001zp}, combining the
string-inspired approach to quantum field theory
\cite{Schubert:2001he} with Monte Carlo methods. As a main advantage,
the worldline algorithm can be formulated for arbitrary Casimir
geometries, resulting in a numerical estimate of the exact answer
\cite{Gies:2003cv}. Since the approach is based on Feynman
path-integral techniques, the problem of determining the Casimir
fluctuation spectrum is circumvented \cite{Gies:2005ym}. The
resulting algorithms are trivially scalable, and computational efforts
increase only linearly with the parameters of the numerics.

Recent results obtained by worldline numerics \cite{Gies:2006bt} go
hand in hand with those obtained by new analytical methods
\cite{Bulgac:2005ku,Emig:2006uh,Bordag:2006vc} which are based on
advanced scattering-theory techniques; excellent agreement has been
found for the experimentally important sphere-plate and cylinder-plate
Casimir configurations.

In the present work, we use worldline numerics to examine Casimir edge
effects induced by a fluctuating scalar field, obeying Dirichlet
boundary conditions (``Dirichlet scalar'').  We compute Casimir
interaction energies and forces between rigid surfaces.  Our results
can directly be applied to Casimir configurations in ultracold-gas
systems \cite{Roberts:2005} where massless scalar
fluctuations exist near the phase transition. For
Casimir configurations probing the electromagnetic fluctuation field,
the results for the universal coefficients may quantitatively differ,
but our values can be used for an order-of-magnitude estimate of the
error induced by edges of a finite configuration, thus providing an
important ingredient for the data analysis of future experiments.

In addition to being a simple and reliable quantitative method, the
worldline formalism also offers an intuitive picture of
quantum-fluctuation phenomena. The fluctuations are mapped onto closed
Gau\ss ian random paths (worldlines) which represent the spacetime
trajectories of virtual loop processes. The Casimir interaction energy
between two surfaces can thus be obtained by identifying all
worldlines that intersect both surfaces. These worldlines correspond
to fluctuations that would violate the boundary conditions; their
removal from the ensemble of all possible fluctuations thereby
contributes to the (negative) Casimir interaction energy. The latter
measures only that part of the energy that contributes to the force
between rigid surfaces; possibly divergent self-energies of the single
surfaces \cite{Graham:2002xq} are already removed. 

For a massless Dirichlet scalar, the worldline representation of the
Casimir interaction energy reads
\cite{Gies:2003cv,Gies:2005ym}
\begin{equation}
E=-\frac{1}{2} \frac{1}{(4\pi)^2} \int_{0}^\infty
\frac{d T}{T^3}\,\left\langle\Theta_\Sigma[\mathbf x]
\right\rangle_{\mathbf x} . \label{eq:ECasW} 
\end{equation}
The expectation value in \eqref{eq:ECasW} has to be taken with respect
to an ensemble of closed worldlines,
\begin{equation}
\langle \dots \rangle_{\mathbf x} := \int_{\mathbf x(T)=\mathbf x(0)}
 \mathcal D \mathbf x \, \dots
e^{-\frac{1}{4} \int_0^T d \tau {\dot{\mathbf x}}^2},
\label{eq:VEV2}
\end{equation}
with implicit normalization $\langle 1 \rangle_{\mathbf x}=1$. In
Eq.~\eqref{eq:ECasW}, $\Theta_\Sigma[\mathbf x]=1$ if a worldline
$\mathbf x$ intersects both surfaces $\Sigma=\Sigma_1+\Sigma_2$, and
$\Theta_\Sigma[\mathbf x]=0$ otherwise. The worldline integral can
also be evaluated locally, e.g., with the restriction to worldlines
with a common center of mass, $\mathbf x_{\text{CM}}$, resulting in
the interaction energy density $\varepsilon(\mathbf x_{\text{CM}})$,
$E=\int d^3x \varepsilon(\mathbf x_{\text{CM}})$. The interaction
energy serves as a potential for the Casimir force between rigid
surfaces; the force is thus obtained by simple differentiation with
respect to the distance parameters. 

The worldline numerical algorithm corresponds to a Monte Carlo
evaluation of the path integral of Eq.~\eqref{eq:VEV2} with a
discretized propertime $\tau$. In this work, we exploit the recent
algorithmic developments detailed in \cite{Gies:2006cq}.

\section{Casimir edge configurations}

\subsection{Perpendicular Plates}

\begin{figure}
\includegraphics[scale=.7]{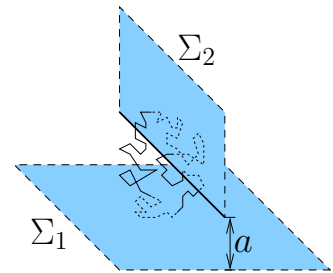}
\caption{Sketch of the perpendicular-plates configuration. The
  minimal distance $a$ between the edge of the upper semi-infinite
  plate (thick solid line) and the lower infinite plate represents the
  only dimensionful length scale in the problem.}
\label{fig:pp}
\end{figure}

\begin{figure}
\includegraphics[width=\columnwidth]{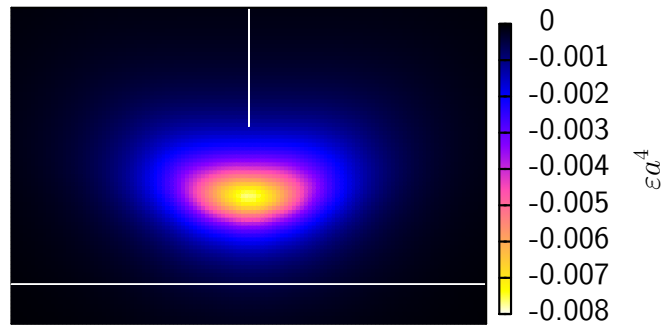}
\caption{Contour plot of the Casimir interaction energy
density $\varepsilon$ for the perpendicular-plate
configuration. The white lines mark the position of the plates to
guide the eye. Ensemble parameters: 2000 loops
with 10~000 ppl.}
\label{fig:contourpp}
\end{figure}

Let us first analyze a semi-infinite
plate perpendicularly above an infinite plate at a minimal distance
$a$, as first proposed in \cite{Gies:2005ym}. This configuration is
illustrated in Fig.~\ref{fig:pp} together with a worldline which
contributes to the Casimir interaction energy, since it intersects
both plates.
This configuration is translationally invariant only in the
direction pointing along the edge with $a$ being the only dimensionful
length scale. The Casimir force per unit length $L$ along the edge
direction is thus unambiguously fixed by dimensional analysis,
\begin{equation}
\frac{F_\bot}{L}=-\gamma_{\bot}\,
\frac{\hbar c}{a^3}.
\label{eq:Ebot}
\end{equation}
Evaluating the worldline integral as outlined above, we obtain an
estimate for the corresponding Casimir interaction energy density
$\varepsilon(\mathbf{x})$, a contour plot of which is given in
Fig.~\ref{fig:contourpp}. For the universal coefficient, we obtain
\begin{equation}
\gamma_{\bot}=1.200(4)\times 10^{-2}.
\end{equation}
The error is below the 1\% level for a path ensemble of 40~000 loops
with 200~000 points per loop (ppl) each. This coefficient is in
agreement with the Casimir interaction energy computed in
\cite{Gies:2005ym}.
\bigskip

\subsection{Semi-infinite plate parallel to an infinite plate}

\begin{figure}
\includegraphics[scale=.7]{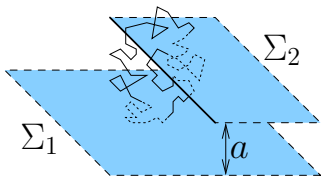}
\caption{Sketch of the configuration of a semi-infinite plate parallel
  to an infinite plate at a distance $a$. A worldline can intersect
  both plates even if its center of mass is located outside the two
  plates.}
\label{fig:hpp}
\end{figure}

Next we consider a first variant of the parallel-plate configuration,
where one of the plates is only semi-infinite with an edge on one
side; see Fig.~\ref{fig:hpp}. This configuration can be viewed as an
idealized limit of a real experimental situation where a smaller
controllable finite plate is kept parallel above a larger fixed
substrate.
In this case, the dominant contribution to the force is given by the
universal classic parallel-plate result of Eq.~\eqref{eq:Casimir} with
$A$ being the surface area of the smaller plate.

\begin{figure}
\includegraphics[width=\columnwidth]{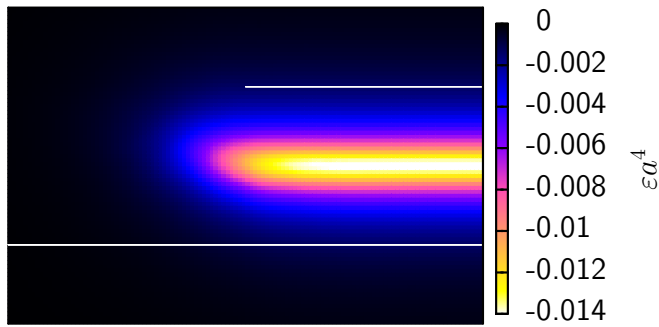}
\caption{Contour plot of the Casimir interaction energy
density $\varepsilon$ for a semi-infinite plate
parallel to an infinite plate. The white lines mark the position of
the plates to guide the eye. The energy-density peak extends into the
outside region, since worldlines can intersect both plates even if
their center of mass is in the outside region. Ensemble parameters:
1000 loops, 10~000 ppl.}
\label{fig:contourphp}
\end{figure}

In the ideal limit of $A$ as well as the edge length $L$ going to
infinity, the sub-leading Casimir edge effect is also
universal. Dimensional analysis requires the exact force to be of the
form
\begin{equation}
F=F_\|-\gamma_\mathrm{1si}\, \frac{\hbar c}{a^3}\, L,\label{eq:F1si}
\end{equation}
where $F_\|$ denotes the parallel-plate force for the Dirichlet
scalar, i.e., without the factor 2 in Eq.~\eqref{eq:Casimir}.  A
priori, the universal coefficient $\gamma_\mathrm{1si}$ can be positive or
negative. The sign can easily be guessed within the worldline
picture: owing to their spatial extent, a sizable fraction of
worldlines can intersect both plates even if their center of mass is
located outside the plates. This can quantitatively be verified by the
energy density, the peak of which indeed extends into the outside
region; see Fig.~\ref{fig:contourphp}.
This peak in the outside region contributes to the total
interaction energy, implying an increase of the Casimir force compared
to the pure parallel-plate formula. Therefore, the Casimir edge effect
leads to further attraction, and the sign of the universal coefficient
$\gamma_\mathrm{1si}$ must be positive. Quantitatively, we find
\begin{equation}
\gamma_\mathrm{1si}=5.23(2)\times10^{-3},\label{eq:1si}
\end{equation}
again with 40~000 loops, 200~000 ppl.\bigskip

\subsection{Parallel semi-infinite plates}

\begin{figure}
\includegraphics[scale=.7]{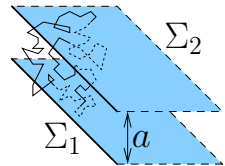}
\caption{Sketch of the configuration with two parallel semi-infinite
  plates at a distance $a$.}
\label{fig:hphp}
\end{figure}

\begin{figure}
\includegraphics[width=\columnwidth]{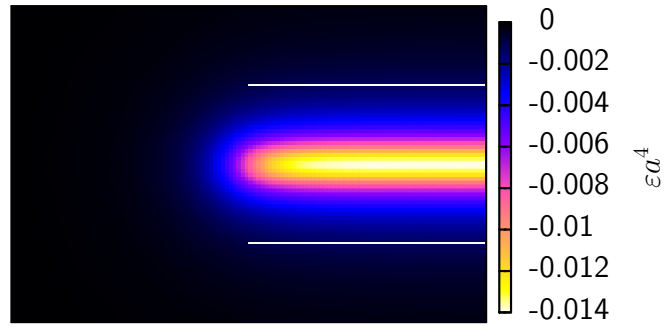}
\caption{Contour plot of the Casimir interaction energy
density $\varepsilon$ for two parallel semi-infinite
plates. The energy-density peak extends into the outside region, since
worldlines can intersect both plates even if their center of mass is
in the outside region. Ensemble parameters: 2000 loops, 10~000
ppl.} 
\label{fig:contourhphp}
\end{figure}

Another variant of the parallel-plate configuration is given by two
parallel semi-infinite plates with parallel edges; see
Fig.~\ref{fig:hphp}. This configuration corresponds to an idealized
parallel-plate experiment where both plates have the same area size
$A$.
In the ideal limit of infinite $A$ as well as infinite edge length
$L$, the exact form of the force is again given by dimensional
analysis,
\begin{equation}
F=F_\|-\gamma_\mathrm{2si}\, \frac{\hbar c}{a^3}\, L,
\end{equation}
equivalent to Eq.~\eqref{eq:F1si}. Qualitatively, the situation is
similar to the preceding one with one semi-infinite
plate. Quantitatively, fewer worldlines in the outside as well as the
inside region near the edge intersect both plates. Both aspects are
visible in the plot of the interaction energy density in
Fig.~\ref{fig:contourhphp}: the peak height and width is reduced near
the edge both inside and outside the plates.
We still observe a positive
universal coefficient, 
\begin{equation}
\gamma_\mathrm{2si}=2.30(1)\times10^{-3}\label{eq:2si}
\end{equation}
(93~000 loops, 500~000 ppl), which is a bit less than half as big as
the preceding case with one semi-infinite plate. Again, the Casimir
edge effect increases the force in comparison with the pure
parallel-plate estimate $F_\|$.

\section{Edge-configuration estimates}

The universal results for the idealized configurations presented
above can immediately be used to derive estimated predictions for
further Casimir configurations. 
\bigskip

\subsection{Casimir comb}

Replicating the perpendicular-plate configuration in the horizontal
direction of Figs.~\ref{fig:pp} and \ref{fig:contourpp}, we obtain a
stack of semi-infinite plates (a ``Casimir comb'') perpendicularly
above an infinite plate. Let $d$ be the distance between two
neighboring semi-infinite plates, i.e., the distance between two teeth
of the comb. In the limit $d\gg a$, we obtain the Casimir force
between the Casimir comb and the infinite plate by simply adding the
forces for the individual perpendicular plates. The reliability of
this approximation is obvious from Fig.~\ref{fig:contourpp}, which
shows that the dominant contribution to the energy is peaked inside a
region with length scale $\sim a$. The resulting force is
\begin{equation}
F_{\text{comb}}=-\gamma_\bot\, \frac{\hbar c}{a^3 d}\, A, 
\end{equation}
with $A=L n d$ being the total area of a comb with $n$ teeth. For a
fixed comb, i.e., fixed $d$, the short-distance Casimir force thus has
a weaker dependence on $a$ than for the parallel-plate case. In the
opposite limit $d\ll a$, we expect the force between the comb and
the plate to rapidly approach that of the parallel-plate case
\eqref{eq:Casimir}. This is because a generic worldline contributing
to the force will have a spatial extent of order $a$, such that the
finer comb scale $d\ll a$ will not be resolved by the worldline
ensemble to first approximation. A similar observation has been made
in studies of periodic corrugations \cite{Emig:2001dx}.
\bigskip

\subsection{Finite parallel-plate configurations}

In a real parallel-plate experiment, the finite extent of the plates
induces edge effects. If the typical length scale $L$ of a plate (such
as the edge length of a square plate or the radius of a circular disc)
is much larger than the plate distance $a$, our results for the
idealized limits studied above can be used within a good
approximation. The force law can then be summarized as
\begin{equation}
F=-\gamma_\|\, \frac{\hbar c}{a^4}\, A_{\text{eff}},
\end{equation}
where the effective area $A_{\text{eff}}$ also carries the information
about the edge effects. For the case of a smaller plate with area $A$
and circumference $C$ above a much larger substrate, the effective
area is given by
\begin{equation}
A_{\text{eff}}=A+\frac{\gamma_\mathrm{1si}}{\gamma_\|}\, aC, \label{eq:A1si}
\end{equation}
e.g., $C=4L$ for a square plate with edge length $L$. For the case of
two parallel plates of equal size and shape with area $A$ and
circumference $C$, Eq.~\eqref{eq:A1si} holds with $\gamma_\mathrm{1si}$
replaced by $\gamma_\mathrm{2si}$. Obviously, the effective area
$A_{\text{eff}}$ is larger than the physical area in either case. 

Consider, for instance, a square plate of edge length $L$ above a
larger substrate: the Casimir edge effects induce a correction on the
1\% level if $a\gtrsim 1$\% of $L$.
In the experiment of reference \cite{Bressi:2002fr}, the  edge length is
$L=1.2\mathrm{mm}$ and the distance goes up to $a=3\mu\mathrm m$. One of
the edges faces an edge of the substrate, similar to Fig. \ref{fig:hphp},
whereas the other three correspond to Fig. \ref{fig:hpp}.
For the Dirichlet scalar this results in a correction of 0.2\%,
which is much smaller than the 15\% precision level of the experiment.

\section{Conclusions}

We have performed a detailed quantitative study of Casimir edge
effects induced by a fluctuating scalar field obeying Dirichlet
boundary conditions. All of our results exhibit a uniquely fixed
dependence on dimensionful scales, as for Casimir's classic
result. The effect of quantum fluctuations is quantitatively encoded
in a universal dimensionless coefficient, which only depends on the
geometry, the nature of the fluctuating field and the boundary
conditions. From the perspective of a scattering-theory approach,
Casimir edge effects are dominated by diffractive contributions to the
correlation functions which are difficult to handle for direct
approximation techniques \cite{semicl,Scardicchio:2004fy}; hence, our
results give an important first insight into the properties of
diffractive contributions to Casimir forces.  For Casimir measurements
involving electromagnetic fluctuations, our results serve as a first
order-of-magnitude estimate of the error induced by edges of finite
configurations -- an error that any parallel-plate experiment has to
deal with.

The authors acknowledge support by the DFG Gi 328/1-3 (Emmy-Noether
program) and Gi 328/3-2.

\end{document}